# Data-Driven Prediction and Evaluation on Future Impact of Energy Transition Policies in Smart Regions

Chunmeng Yang[1], Siqi Bu[1], Yi Fan[2*], Wayne Xinwei Wan[3], Ruoheng Wang[1], Aoife Foley[4]

## Abstract

To meet widely recognised carbon neutrality targets, over the last decade metropolitan regions around the world have implemented policies to promote the generation and use of sustainable energy. Nevertheless, there is an availability gap in formulating and evaluating these policies in a timely manner, since sustainable energy capacity and generation are dynamically determined by various factors along dimensions based on local economic prosperity and societal green ambitions. We develop a novel data-driven platform to predict and evaluate energy transition policies by applying an artificial neural network and a technology diffusion model. Using Singapore, London, and California as case studies of metropolitan regions at distinctive stages of energy transition, we show that in addition to forecasting renewable energy generation and capacity, the platform is particularly powerful in formulating future policy scenarios. We recommend global application of the proposed methodology to future sustainable energy transition in smart regions.

*Keywords*—Energy transition, Renewable energy, Policy prediction, Policy evaluation, Machine learning

[1]Department of Electrical Engineering, Hong Kong Polytechnic University. [2]Department of Real Estate, National University of Singapore. [3]Department of Banking and Finance, Monash University. [4]School of Mechanical and Aerospace Engineering, Queen's University Belfast. * indicates corresponding author. E-mails: yi.fan@nus.edu.sg.

# 1. Introduction

## 1.1. Background

Cities are large and complex hubs of energy consumption [1]. To meet widely recognised carbon neutrality targets, renewable energy systems are being established in metropolitan areas to improve sustainability in the context of fossil fuel depletion and climate change [2,3]. Recent decades have witnessed remarkable developments in renewable energy technologies; according to the 2019 Renewables Global Status Report of United Nations Environmental Programme (UNEP), renewable energy now constitutes around 10.6% of total global energy consumption.

However, the rapid implementation of renewable energy technologies in numerous electricity grids presents challenges, such as major upgrades of existing infrastructure, the decline of established business models, utility companies' intensified economic struggles, and uncertainty regarding the performance of the electricity sector after integrating renewables [4,5]. Energy transition policies have been enacted to ensure a stable and economical transition from fossil fuels to clean energy and promote the sustainable adoption of new technologies in energy markets [6].

The goal of energy transition policies is to fulfil regional energy demand and sustain economic growth while meeting targets for environmental protection. Such policies are expected to be evaluated and adjusted in a timely manner in response to dynamic changes in economic and environmental development, and to incorporate unexpected shocks such as climate change. However, this can be difficult for policymakers, because they lack a comprehensive tool for accurately predicting and quantitatively evaluating the future impact of energy transition policies in complex urban contexts and conducting effective forecasting. Most existing policy evaluations consist of policymakers' technical judgements based on retrospective experience [5,7–9], which is less efficient. Therefore, there is a pressing need to develop an effective tool for the impact prediction and evaluation of energy transition policies.

## 1.2. Literature Review

Effective future impact prediction and evaluation of energy policies is always a challenging problem due to uncountable factors including local natural resources, overall demand for energy driven by economic growth, human behaviours and societal development, and energy market structure, mechanism and efficiency [10–12]. Traditional statistical and qualitative analytic models have been used for forecasting [13–15], but these methods require strong ex ante assumptions and are less flexible for complex urban energy systems. Also, there is a lack of available platform for policy evaluation implementation and validation.

Existing literature adopts different methods for approximating energy policy evaluation with different research focuses. Some simulation-based methods employ market simulation, dynamic modelling, and multi-agent models [16], etc. These methods normally require highly detailed information of market operations or energy policies for modelling and may entail many pre-determined assumptions to ease problems. Based on a reliable simulation, some optimisation algorithms or evolutionary game theory can be applied to compute the equilibriums of the specific market dynamics under the impact of the designated policies. Some other methods may adopt economics related techniques to estimate the policy impact, such as macro-economic models and input/output models [17], and they are generally economic experience-based methods. Some qualitative methods for energy policy evaluation are generally theory based, such as multi-criteria analysis [18] and policy theory analysis [19]. Whereas some of these methods are heavily affected by data availability and reliability, its nature of theoretical analysis also requires intensive logic analysis and empirical inference. Some other efforts are devoted to provide post-treatment analysis of the impact on economic growth and energy generation [20–22]. Hence, it can be revealed from the review above that most of policy evaluations mainly focus on the current or short-term impact of the energy policies. There is a lack of research on the prediction and evaluation on the long-term impact of the energy policies to facilitate the energy policy planning and adjustment.

The technology diffusion model has been widely used to evaluate the evolutionary process of the policy future impact, and empirically explain the dynamic energy transition in power generation and consumption due to the introduction of renewable energy technologies [23–26]. The simulated process of policy impact can be classified into three phases: 1) slowly developing for a long period; 2) rapidly taking off with a high technological impact; 3) gradually saturating after a certain time in the future [16]. In [27], an agent-based model incorporating the bass diffusion model is developed to model the diffusion process of renewable energy technologies in consideration of the effects of disruptive innovation. A general technology diffusion theory is adopted in [28] to forecast different types of renewables including wind power, solar power, and biomass power, and proposes a renewable energy optimisation model using diffusion results as inputs to calculate the maximum size of renewable generation in the future. In a related literature [29], the spatiotemporal diffusion model is used to forecast the capacity of distributed energy resources (DER), where a temporal module projects the global and local diffusion process of DERs and a cellular model is employed to calculate the correlations between regions using the innovativeness score mechanism. Thus combining with the diffusion model, we can predict and evaluate the long-term impact of new energy policies based on the "business-as-usual" scenario for the energy policy planning and adjustment.

### 1.3. Contributions and Paper Organization

To effectively address all the points above, a novel data-driven platform that combines an artificial neural network (ANN) model and a technology diffusion model is developed in this paper to predict and evaluate the future impact of energy transition policies. Compared with most policy evaluation methods tackling short-term evaluation problems, the proposed method is a combination of the empirical analysis with policy information collected for supporting an effective policy scoring mechanism, and a data-driven quantitative analysis to precisely forecast near and far future scenarios based on multi-source data, so that the gap between the projected and targeted performances can be minimised.

The effectiveness of the proposed platform is demonstrated by using case studies in three smart metropolitan regions with different energy transition stages: Singapore, London, and California. By collecting monthly information on macroeconomic conditions, electricity market structures, and current renewable energy development, future trends in renewable energy transition up to 2025 are predicted in the three regions under different scenarios: a baseline scenario without any new policy incentives to promote energy transition, a theoretical scenario that illustrates the optimal process of renewable energy transition, and a predicted scenario based on actual new energy transition policies announced in each region. Based on the predicted results of the platform, policy recommendations on the sustainable development of renewable energy are provided for government agencies and other stakeholders. Our proposed methodology marks an advancement in the frameworks used to qualitatively evaluate energy transition, which mainly relies on the empirical predictions of policymakers and requires frequent adjustments [5]. The proposed platform can be generalised to other metropolitan regions to predict and evaluate the future impact of energy transition policies. It will also facilitate communications and decision-making among disparate stakeholders. The major contributions of the paper are summarised below:

1. We make a methodological contribution by developing a novel platform for renewable energy transition policy evaluation and forecasting. The platform can not only effectively forecast and evaluates the future impact of existing and new energy policies at different development stages, but also be adjusted in a timely manner in response to dynamic changes in economic and environmental development, as well as incorporating unexpected shocks such as financial crisis or pandemic. Specifically, we combine the machine learning artificial neural network (ANN) model with the technology diffusion model, which enables forecasting of dynamic pathway to achieve future policy targets by feeding proposed policies and provides a possible interface to energy policy adjustment utilizing the projected policy evaluation results.

2. To test the effectiveness of the platform, we make the data contribution by collecting a rich set of determinants of renewable energy development from three smart regions at different development stages. Specifically, we collect a comprehensive set of renewable energy policies in Singapore, London, and California, and apply policy scoring mechanism to each policy to evaluate effectiveness. In addition, we collect monthly data between 2004 and 2019 on a region's level of economic prosperity (i.e., GDP, interest rate, population, market prices of renewable energy and conventional fossil energy substitutes; operational modes in different stages of energy production), availability of natural resources (i.e., sunshine duration, wind), green ambitions including relevant technology advancements and human capital, society's culture and governments' commitment. We also collect monthly electricity production from renewable sources and renewable power generation capacity in each smart region, as outcome variables.

3. Most existing policy evaluations consist of policymakers' technical judgements

based on retrospective experience, which can be inefficient or even biased, and generally tackle short-term evaluation problems. By feeding the rich policy, socioeconomic, environment, and energy data into the novel data-driven platform, we contribute to accurately predict and quantitatively evaluate the future impact of energy transition policies in complex urban contexts. Moreover, this platform is powerful in formulating future policy scenarios and forecasting the development pathway of renewable energy production and capacity. We thus recommend application of this proposed methodology to future sustainable energy transition in other smart regions.

The remainder of this paper is organised as follows. The data-driven policy prediction and evaluation platform is developed in Section 2. Section 3 presents case studies and results to validate the established platform. Based on the proposed prediction and evaluation of energy policies, the effective adjustment of energy policies is conducted. On this basis, some in-depth discussions and useful recommendations are provided for three smart regions in Section 4. Major conclusions are drawn in Section 5.

## 2. Methodology

### 2.1. Data-driven Policy Prediction and Evaluation Platform

In this paper, we provide a novel platform to predict and evaluate the future impact of energy transition policies. The platform is built on machine learning along different dimensions in a complex urban energy system, and thus provides reliable prediction of renewable power generation capacity. In addition, diverse hypothetical policy scenarios can be compared on the platform, with flexibility in timely adjustment via a feedback loop among factors input, policy instruments, and renewable energy output. We combine two methodologies—machine learning and technology diffusion modelling—and extend their applications to quantitatively forecast regional energy transition progress, evaluate the effectiveness of energy transition policies in a timely manner, and adjust energy transition policies based on preset transition targets. Fig. 1 summarises this data-driven platform's structure. First, a comprehensive set of determinants for local renewable energy developments is collected. Their monthly data are then input into an artificial neural network (ANN) to train a model for the baseline forecast of the region's renewable power generation/capacity. Finally, we investigate the sensitivity of the renewable energy transition rate (i.e., the proportion of renewable energy in total energy consumption) in response to different policy instruments using the technology diffusion model, which provides guidance on policy formation, selection, and optimisation.

Predicted regional renewable energy generation capacity and annual electricity production from renewable sources are the outcomes of our proposed platform. The evolving trajectory of the outcomes under any specific policy instrument lies between the current "business-as-usual" scenario without any new policy impact (i.e., the baseline predictions from the ANN model) and the optimal scenario (e.g., the fastest possible sustainable growth of renewable power generation). Compared with the optimal scenario, the predicted scenario under any policy instruments is determined by two additional model factors: diffusion ceiling ($f_c$) and diffusion speed ($f_p$). These are estimated by comparing the region's actual energy transition policy with the theoretically optimal policy using a policy scoring mechanism [30–33].

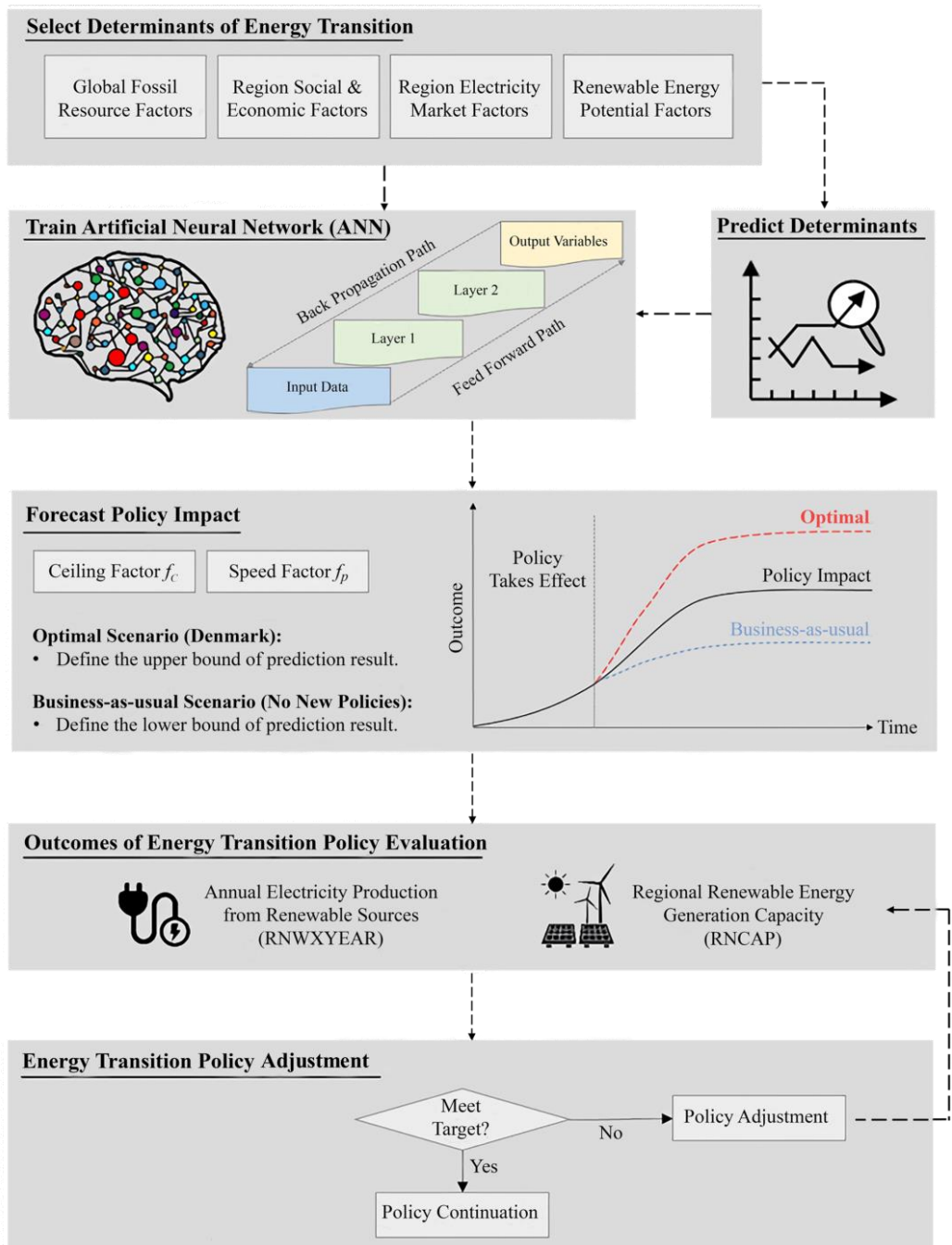

**Fig. 1** Data-driven Energy Transition Policy Prediction and Evaluation Platform Structure

Specifically, we evaluate the intensity of the new energy transition policy (i.e., $f_c$ and $f_p$) in a set of dimensions with different weightings. In each dimension, we first set a theoretically optimal target or benchmark. Then, we calculate scores for the actual policy along each dimension by comparing the policy with the one under the optimal setting, conditional on the pre-policy context. If there is a precise quantitative target for the new policy (e.g., an index that measures the smartness of power grids), we scale the policy's achievement linearly in the range of 0%-100%, with the pre-policy context at 0% and the optimal one at 100%. For other dimensions without precise quantifiable targets for outcome variables, we categorise them into five levels and assign the intensity scores of 0%, 25%, 50%, 75%, or 100%, respectively, for illustrative purpose. The overall intensity factors are the weighted averages of the scores across all dimensions (details are provided in Subsection 2.4).

An important contribution of our policy scoring mechanism is that it captures the dynamic impact of a policy at different energy transition stages. For instance, consider a financial incentives (FI) policy, such as tax reductions and discounted loans (Fig. 2). If this FI policy is initially introduced to a region at a very early stage of energy transition, it is expected to be less effective than the same FI policy applied in a region at a more mature stage of energy transition. This is because the transition cost (e.g., establishing infrastructure and raising awareness of the transition) is much higher at the very early stage. Also, the effectiveness of a typical policy is expected to decline over time. One plausible reason is that governments tend to reduce such subsidies gradually if the FI policy has been in effect for a while during the process of cultivating the market. To summarise, our policy scoring mechanism not only evaluates the absolute intensity of a policy, but also considers its relative effectiveness in specific regions along the temporal dimension.

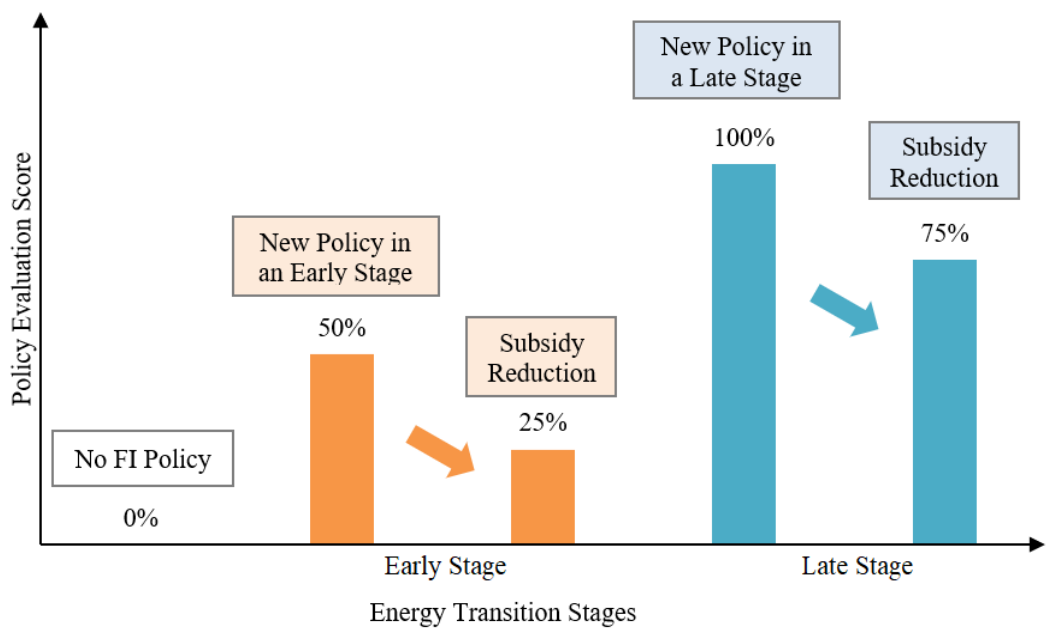

**Fig. 2** Evaluation of an Energy Transition Policy at Different Stages

## 2.2. Determinants Selection

We follow the "Two Degrees" framework of National Grid ESO [34] to select the factors that impact future energy generation in metropolitan areas. First, the region's level of economic prosperity impacts the demand for energy, as well as the capital available for government expenditure, business investment, and consumer spending on renewable energy development. Second, the region's green ambitions indicate the degree of government's engagement with reducing carbon emissions and increasing sustainability, conditional on natural resources and technological restrictions [35].

Along the dimension of economic prosperity, the energy supply must meet the underlying demand for energy consumption to ensure sustainable growth [36,37]. On the demand side, macroeconomic factors—such as GDP, the interest rate [3], and the domestic population—correlate with demand for energy consumption. A significant proportion of the energy required is for electricity, as reflected by its high level of consumption. On the supply side, the production of renewable energy is determined by both the market price of renewable energy and the market price of conventional fossil energy substitutes (e.g., coal, crude oil, natural gas, etc.). In particular, fluctuations in the price of domestic electricity will increase the revenue risk of renewable energy investment, which decreases the enthusiasm of investors in the renewable energy market [10,11]. The cheaper production costs of fossil energy than renewable energy may also lower the production of renewable energy [12]. Lastly, the operational modes in different stages of energy production—generation, transmission, distribution, and retail—also impact the renewable energy supply.

Along the dimension of green ambitions, one of the most influential driving factors is the availability of local natural resources. Regions with longer sunshine duration are normally more motivated to develop solar energy, since solar irradiation is clean and inexhaustible [38]. In contrast, areas with high winds are more likely to use wind turbines [39]. Other key drivers of green ambitions are advancements in technology [40] and the availability of human capital [41,42], which will significantly decrease the levelised cost of generating renewable energy. Lastly, the society's culture; the region's civilisation level, measured by the human development index (HDI); and the government's commitment to sustainable development (e.g., commitment to $CO_2$ emission targets [43]) are also crucial. We do not directly adopt climate change indicators, because it is difficult to quantify them using a specific indicator. Nonetheless, we note that determinants such as wind resources and solar irradiation can, to some extent, reflect climate trends, and indices regarding the degree of government's commitment to climate issues can also potentially involve the impact of climate change.

These determinant factors are collected from open sources in London (2004-2018), California (2004-2019), and Singapore (2009-2019). In the Appendix, we summarise in detail the sources and predictions of these input data. All categorical variables are one-hot encoded and numerical variables are standardised. Future changes in the determinant factors up to 2025 in the three study regions are predicted according to the following rules. First, for factors with available predictions from government agencies (e.g., regional GDP, levelized cost of energy (LCOE) of solar photovoltaic (PV) or wind turbines), we use the official data. Second, for the measurements of nature resources (e.g., solar radiation, wind speed) and local economic policies (e.g., interest rate), we assume that these remain unchanged. Lastly, for other time-variant variables, such as global fossil energy prices, we use the historically average growth rate to predict future values.

## 2.3. The ANN Model

A feed-forward ANN is constructed for the baseline prediction of renewable energy output. Historical data are partitioned into training and validation data sets. The selected determinant factors (i.e., input) are mapped to renewable energy generation/capacity (i.e., output) through a series of intermediate mapping layers (denoted hidden layers), and the model recursively trains the weightings in the hidden layers to minimise prediction errors in the training data. The model is then applied to the validation data set to test the predictive accuracy, which addresses the issue of potential overfitting. Lastly, we form the baseline prediction of future renewable power generation/capacity using the ANN model and the estimated future time series of the selected determinant factors.

Specifically, we employ an ANN model with multiple hidden layers. The number of neurons in the first layer equals the input number of determinants, and the numbers of intermediate parameters (neurons) in each hidden layer are half the size of their previous layers to extract high-level features for prediction. For each layer connection, we use the rectified linear unit function for activations. For loss function, we implement the smooth loss calculation:

$$loss(x, y) = \frac{1}{n} \sum_i z_i, \tag{1}$$

where $x$ and $y$ denote the vectors of predicted and actual values. $z_i$ is given by

$$z_i = \begin{cases} 0.5(x_i - y_i)^2, & if\ |x_i - y_i| < 1 \\ |x_i - y_i| - 0.5, & otherwise \end{cases}. \tag{2}$$

For the optimiser of the weights in the model, the *RMSprop* algorithm [44] is used to keep a shifting average of squared gradient at time *t* for each weight $w$:

$$E[g^2](t) = 0.9 E[g^2](t-1) + 0.1 \left(\frac{\partial Loss(t)}{\partial w(t)}\right)^2, \tag{3}$$

where $E[g^2]$ is the moving average of squared gradients and $\partial Loss/\partial w$ is the gradient of the loss function with respect to the weights. The moving average parameter is set to the conventional default value of 0.9. Then, the weights are updated using the moving average of squared gradients:

$$w(t) = w(t-1) - \frac{\eta}{\sqrt{E[g^2](t)}} \frac{\partial Loss(t)}{\partial w(t)}. \tag{4}$$

$\eta$ denotes the learning rate and is set to 0.001, as the default number in the *RMSprop* algorithm. We randomly sampled 95% of the collected data as the training data set and include the other 5% of the data in the validation data set. The predictive accuracy of our baseline model on the data set is around 95%.

## 2.4. The Technology Diffusion Model

Based on the ANN prediction model derived in Subsection 2.3, the impact of existing energy transition policies can be predicted and evaluated, which is known as the baseline scenario or "business-as-usual" scenario here. In this subsection, the technology diffusion model will be employed to predict and assess the impact of different types of new energy transition policies to be implemented.

In general, the energy transition policies can be classified into direct policies (e.g., obligations and certificates, feed-in-tariffs, and financial incentives), integrating policies, and enabling policies [45]. Direct policies take effects at a strategic level to impact the target renewable energy generation. Particularly, the direct policy can be classified into three main types, namely obligations & certificates, feed-in-tariff, and financial incentives [46]. Obligations refer to the policy instruments that outline the required renewable energy generation, and certificates are provided to acknowledge the compliance with the obligations. The feed-in-tariff policy encourages end-users to generate more renewable energy (e.g., install solar photovoltaic or wind power system at premises), by promising to purchase these energies back at a higher price than the market energy tariff. Financial incentives are normally provided for the large-scale application of new renewable energy technologies. In contrast, integrating policies and enabling policies take effect at the operational level. Specifically, the former provides guidance to the adoption of new renewable energy technologies into the existing energy system and institutional framework, while the latter aims to ensure effective and stable operation of the renewable energy systems.

At an early stage of adopting a new renewable energy technology, the transition policy mainly works by stimulating relevant agencies to fund research institutions, set up demonstration programs, and gradually acquaint residents and other stakeholders with the new renewable energy technology [47,48]. After that, policy instruments are applied to promote on-grid power generation from renewable energy, normally through the provision of financial incentives [49–51]. As the generation cost of renewable energy gradually decreases as a result of economies of scale, policymakers then aim to optimise the local market structure of fossil fuels and renewable energy and balance the interests of diverse stakeholders in order to sustain both regional economy growth and green ambitions [52–54].

The technology diffusion model assumes that after the introduction of a new policy, the renewable power generation capacity ($RNWXYEAR$) and electricity production ($RNCAP$) will follow the s-curve in the equation below:

$$R(t) = \frac{c}{1+e^{-pt}}, \tag{5}$$

where $R(t)$ is the renewable generation capacity or electricity production after the policy takes effect. $c$ is the ceiling of renewable power generation capacity or electricity production in the market, which is the saturation level. $p$ is the speed of diffusion and $t$ is the time since the policy takes effect (measured in years).

In the optimal scenario (i.e., Denmark), the ceilings ($c_{op}$) of renewable power generation capacity and electricity production are assumed to be 80%. The average diffusion speed ($p_{op}$) from 1994 to 2018 in Denmark is directly observable in the market, which equals 0.120 for renewable power generation capacity and 0.162 for electricity production. In the baseline scenario without any additional policy interventions, we assume the ceilings ($c_{base}$) of both the renewable power generation capacity and electricity production to be 15% [55,56], and their diffusion speed ($p_{base}$) to be one-third of the corresponding speed in the optimal scenario.

Further, to model the ceiling and diffusion speed of each new policy $i$, two policy-specific factors are defined: $f_{ci}$ denotes the policy's ceiling factor of saturation and $f_{pi}$ denotes the policy's speed factor of diffusion. They represent the intensity of the new policy compared with the baseline scenario and the optimal scenario:

$$c_i = c_{base} + (c_{op} - c_{base}) \times f_{ci}, \tag{6}$$

$$p_i = p_{base} + (p_{op} - p_{base}) \times f_{pi}. \tag{7}$$

Specifically, the $c_{base}$ and $p_{base}$ are generated from the ANN model as proxy for the business-as-usual scenario, which reflect the relationship between ANN model and technology diffusion model. The policy's ceiling factor ($f_{ci}$) and speed factor of diffusion ($f_{pi}$) are generated through the policy scoring mechanism, which reflect the effects of different new energy policies on the two concerned renewable targets. Based on these parameters, the technology diffusion model is then applied to link the policies with the renewable energy generation and capacity. Examples of our proposed policy scoring mechanism from case studies in Section 3 are presented in Tables 1 and 2 and $f_{ci}$ and $f_{pi}$ are calculated for the three smart regions (i.e., Singapore, London and California) respectively.

**Table 1** Evaluation Matrix of Ceiling Factor $f_c$

| Evaluation Index $I_n$ | Score of Evaluation Index $S_n$ | Score of Region | Weight $w$ |
|---|---|---|---|
| **A. Long-term Ambitious for Renewable Energy (RE) Development** | | | **60%** |
| Long-term (2050) Energy Mixture Transition Ambition | 100: Legally compulsory, ambitious, and concrete energy transition strategy exists<br>75: Legally compulsory, ambitious energy transition strategy exists but lack of concretion<br>50: Energy transition strategy exists but lack of concretion and ambition<br>25: No long-term energy transition plan but mid-term (2030) plan exists<br>0: No energy transition strategy | Singapore:25<br>London:100<br>California:100 | 30% |
| Long-term (2050) Renewable Energy Generation Target | Linearly scaled between 0 to 100<br>100: Vision to 100% RE in 2050<br>0: Equal or less than 1%, or no target | Singapore:43<br>London:50<br>California:100 | 30% |
| **B. Electricity Market Structure and Regulation** | | | **40%** |
| Grid-connection Permission | 100: Grid-connection permitted<br>50: Grid-connection partially permitted<br>0: Grid-connection not permitted | Singapore:100<br>London:100<br>California:100 | 20% |
| Demand-side | 100: Both market design and regulatory | Singapore:50 | 20% |

| Evaluation Index $I_n$ | Score of Evaluation Index $S_n$ | Score of Region | Weight $w$ |
|---|---|---|---|
| Management | management integrated<br>50: Either market design or regulatory management integrated<br>0: Not exist<br>*Note*: Market design refers to the usage of Time-of-use Tariff and/or Smart Meter Billing; Regulatory management refers to the implementation of Energy as a Service (EaaS) | London: 100<br>California:100 | |
| **Result $f_c$** | | Singapore: 0.504<br>London:0.850<br>California:1 | |

**Table 2** Evaluation Indexes of Technology Diffusion Speed Factor $f_p$

| Evaluation Index $I_n$ | Score of Evaluation Index $S_n$ | Score of Region | Weight $w$ |
|---|---|---|---|
| **A. Mid-term Ambitious for Renewable Energy (RE) Development** | | | **20%** |
| Mid-term (2030) Renewable Energy Generation Target | Linearly scaled between 0 to 100<br>100: Vision to 100% RE in 2030<br>0: Equal or less than 1%, or no target | Singapore:8<br>London:30<br>California:50 | 10% |
| Dirty Energy Elimination | 100: Dirty fossil fuels phasing out strategy fixed and existed<br>50: Dirty fossil fuels phasing out strategy under discussion but not fixed<br>0: Dirty fossil fuels phasing out strategy does not exist | Singapore:100<br>London:100<br>California:100 | 10% |
| **B. Adoption of Renewable Energy (RE) Policy** | | | **60%** |
| Adoption of Obligations and Certificates Policy (Direct Policy) | 100: Adoption of Obligations and Certificates (O&C) Policy with relaxing regulation trend<br>75: Adoption of O&C Policy without relaxing regulation trend<br>50: Adoption of premilinary O&C Policy with relaxing regulation trend<br>25: Adoption of preliminary O&C Policy without relaxing regulation trend<br>0: No O&C Policy adopted<br>*Note*: Major O&C Policy includes Renewable Portfolio Standard (RPS) (e.g., RPS in USA), Renewable Purchase Obligations (RPO) (e.g., RPO in India), Renewable Electricity Certificates, etc. | Singapore:0<br>London:75<br>California:100 | 18% |
| Adoption of Feed-in Tariff Policy (Direct Policy) | 100: Adoption of Feed-in Tariff (FiT) Policy without foreseen tariff reduction inclination<br>75: Adoption of FiT Policy with tariff reduction inclination<br>50: Adoption of initial FiT Policy without foreseen tariff reduction inclination<br>25: Adoption of initial FiT Policy with tariff reduction inclination<br>0: No FiT adopted<br>*Note*: Major FiT Policy include Feed-in-Tariff, Feed-in Premium (FIP), etc. | Singapore:50<br>London:75<br>California:75 | 18% |
| Adoption of Financial Incentives | 100: Adoption of Financial Incentives (FI) Policy without foreseen subsidy | Singapore:50<br>London:75 | 14% |
| Policy (Direct Policy) | reduction inclination<br>75: Adoption of FI Policy with subsidy reduction inclination<br>50: Adoption of initial FI Policy without foreseen subsidy reduction inclination<br>25: Adoption of initial FI Policy with subsidy reduction inclination<br>0: No FI Policy adopted<br>*Note*: Major FI Policy includes tax incentives, attractive loans, etc. | California:75 | |
| Adoption of Integrating Policy | Supportive policies to develop or upgrade grid codes, adjust system operations, or expand power grid for integration of renewable energy:<br>100: Adoption of mature Integrating Policy<br>50: Adoption of preliminary Integrating Policy<br>0: Not exist<br>*Note*: Major Integrating Policy includes Enhanced Frequency Response (EFR), grid access registration, etc. | Singapore:50<br>London:100<br>California:100 | 5% |
| Adoption of Enabling Policy | Supportive policies to enable the implementation and highlight the benefits of renewable energy without additional cost to government or customers:<br>100: Adoption of mature Enabling Policy<br>50: Adoption of preliminary Enabling Policy<br>0: Not exist<br>*Note*: Major Enabling Policy includes National Climate Change Plan, Climate Protection Policy, etc. | Singapore:100<br>London:100<br>California:100 | 5% |
| **C. Electricity Market Structure and Climate Change Performance** | | | **20%** |
| Smart Grid Index (SGI) | A quantifiable framework to measure the smartness of regional power grid considering supply reliability, renewable energy integration, network security, etc.<br>*Note*: Score scale is between 0 to 100 and refers to the 2019 SGI benchmarking results. | Singapore: 66<br>London: 89<br>California: 93 | 10% |
| Climate Change Performance | A quantifiable framework to evaluate national climate policy performance including greenhouse gas emission and climate policy<br>*Note*: Score scale is between 0 to 100 and refers to the Climate Change Performance Index 2020. For Singapore, which is lack of data, it is assumed as the same performance level of China. | Singapore:48.16<br>London: 69.80<br>California:18.60 | 10% |
| **Result $f_p$** | | Singapore: 0.527<br>London: 0.764<br>California: 0.782 | |

## 3. Case Studies and Results

We demonstrate our data-driven energy transition policy evaluation platform in three metropolitan cities/regions: Singapore, London, and California. These cities/regions were selected as case studies because they all have developed economies with significant energy demand, but are at distinctly different stages of renewable energy transitions. Depending on the geographic variations in the types and quantities of renewable energy resources and the levels of policymakers' commitment to carbon neutrality, we classify Singapore, London, and California as case studies at the primary, intermediate, and advanced stages of renewable energy transition, respectively.

Singapore is at a primary stage of renewable energy transition. While Singapore holds a strong ambition to develop clean energy as London and California, it is not expected to aggressively increase the percentage of renewable energy in its total energy consumption due to the limit of renewable resources. Solar power is one of the few available renewable sources in Singapore, but it is challenging to install photovoltaic panels in a densely populated city-state with insufficient land. The current strategy of the Singapore government is to focus on the research and development in improving the power generation efficiency within limited areas.

London is an example of a metropolitan area which is in an intermediate energy transition stage and is expected to have substantial improvements in renewable energy development in near future. UK is one of the best locations for wind power generation in Europe. Its wind farms are developed rapidly and have produced more electricity than coal power plants since 2016. The solar energy generation in the UK has also increased rapidly in recent years, due to the government's feed-in tariff schemes and the technology advancements in reducing the cost of photovoltaic panels. The London government aims to supply 15% of London's energy by renewable energy by 2030.

California is in an advanced position of renewable energy transition worldwide, with a long history of renewable energy development since 1990s. Up to 2020, renewable electricity generation constitutes around 50% of the net electricity generation in California, and nuclear and hydroelectric power are the major renewable resources. Looking ahead, the California government maintains its advancement in green ambition but has progressively shifted its focus of renewable transition from expanding the overall renewable generation capacity to developing new renewable energy resources.

The online supplementary materials on "Review of Energy Transition Policies in Singapore, London, and California" describe the renewable energy development and policies in the three regions over recent decades [57]. To represent the policy impact on energy transition, as described in the platform description, we focus on two outcomes of prediction: annual electricity production from renewable resources ($RNWXYEAR$) and renewable power generation capacity ($RNCAP$). We conduct a short- to medium-term analysis up to 2025.

## 3.1. Determinants of Regional Energy Transition

Due to the complexity of renewable energy development at the regional level, selecting appropriate determinant factors is crucial for reliable estimation of renewable power generation and capacity. According to the National Grid ESO (which operates the Great Britain's electricity system), a city's future scenario of renewable energy utility is influenced by its economic prosperity and societal green ambitions [34], as shown in Fig. 3. The prosperity of the economy impacts the demand for energy, as well as the capital available in return for government expenditure, business investment, and consumer spending in renewable energy development. The green ambition indicates the level of government's engagement with reducing carbon emissions and increasing sustainability, conditional on the natural resources and technological restrictions [58].

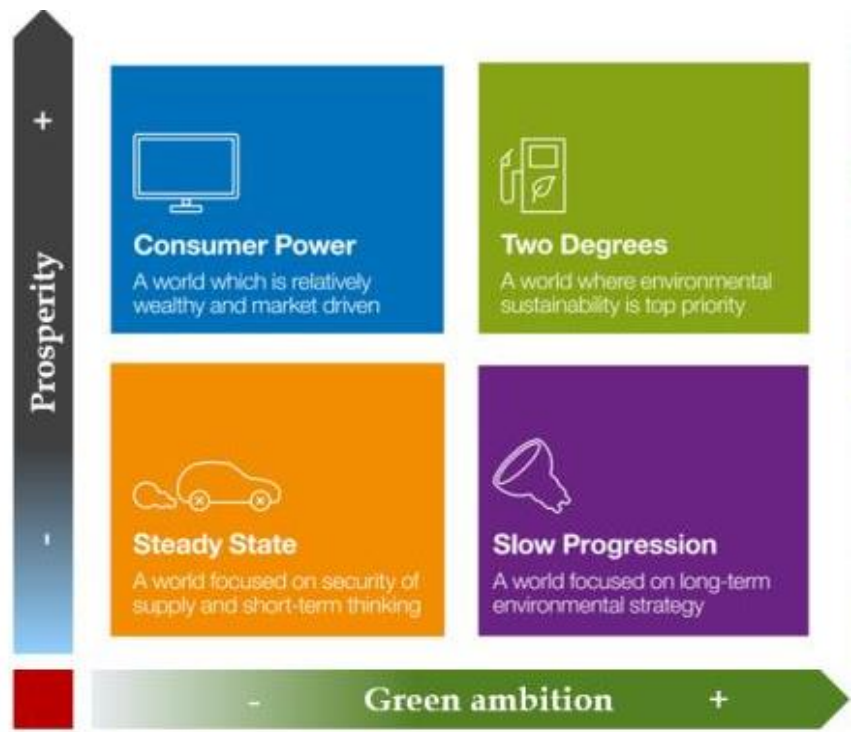

**Fig. 3** The Future Energy Scenarios. Data source: National Grid ESO of the UK.

We follow this future energy scenarios to screen the determinants of regional energy transition and use the selected factors as the input in our proposed renewable energy policy forecasting platform. Fig. 4 presents our analytic framework. Along the dimension of economic prosperity, influential factors include the GDP, interest rate, domestic population, total electricity consumption, market price of renewable energy and its fossil energy substitute, and local operation model of energy production. Along the dimension of societal green ambitions, we select influential factors: the availability of natural resources, human development index (HDI), and human capital and technology advancements in the region, as well as the local culture and the government's commitment to sustainable development. Then we run a correlation analysis to exclude non-representative determinants and select the most influential factors in our study areas. For instance, Singapore is located near the equator, with limited wind power resources, so wind only weakly correlates with renewable power generation/capacity; thus, we eliminate the corresponding factors from our model for Singapore.

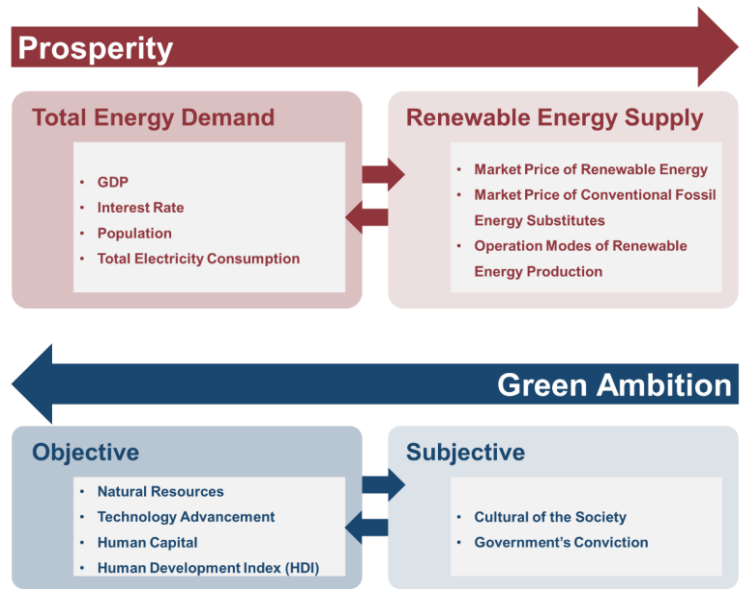

**Fig. 4** The Proposed Analytic Framework for Determinant Selection

### 3.2. Policy Impact Prediction and Evaluation

With reasonably selected determinants for the different cases of Singapore, London, and California, the datasets consist of the sample pairs of determinants and target indices—RNWXYEAR and RNCAP—which are constructed for further training of the ANN-based prediction models. The input size of the model varies with the number of determinants, which are concatenated as a single input vector. Before feeding the determinants into the model, the raw data are preprocessed by standardising the scales of the determinants with their means and standard deviations. The final input size of the datasets for Singapore, London, and California are 12, 14, and 14, respectively. Since limiting the complexity of neural networks helps alleviate the overfitting problem for small datasets, two hidden layers with neurons 12 and 6 are constructed for feature learning. The training epoch is set at 5,000 for early stopping, and mini-batch size is set at 32. We augment the datasets to ease the small-sample learning problem by using the extra information from the monthly time resolution of determinants: The monthly changing data on determinants with corresponding interpolated labels of annual RNWXYEAR and RNCAP are employed to smooth the learned curves of historical renewable energy data, thereby gaining a good variance-bias trade-off and avoiding possible overfitting on training sets.

We first evaluate the feasibility of the proposed prediction method with the determinants by using the mean absolute percentage error (MAPE) to compare the normalised training and test errors. The MAPE is given as

$$MAPE = \frac{1}{N}\sum_{i}^{N} \frac{|P_i^R - P_i^P|}{P_i^R} \times 100\%, \tag{7}$$

where $N$ is the number of points to be evaluated, $P_i^R$ is the $i^{\text{th}}$ data point in reality, and $P_i^P$ is the $i^{\text{th}}$ data point predicted by the model. The MAPE measures the predictive accuracy via the percentage of errors to the true values. The results, averaged over 100 repetitive experiments, for datasets from Singapore, London, and California are shown in Table 3, in which data for the latest 2 years are used for the test set and the remaining data for training, given the small size of the entire datasets. The model is tested on the Singapore dataset, which has the largest error levels of the three datasets; this is probably because the patterns between determinants and two forecasted indices are less correlated. Even with the small data size, the model still reaches a decent accuracy level, with a MAPE near 12%. Results for London and California indicate significantly better fitting for the RNWXYEAR and RNCAP, with MAPEs less than 2% and 3.5% for the training and test sets, respectively. In all, the tests demonstrate a high performance of the model, at the same time verifying the feasibility of proposed prediction scheme.

**Table 3** Performance of renewable energy prediction on training and test data sets

| MAPE (%) | Singapore | | London | | California | |
|---|---|---|---|---|---|---|
| | RNWX-YEAR | RNCAP | RNWX-YEAR | RNCAP | RNWX-YEAR | RNCAP |
| Training | 6.95 | 9.85 | 1.94 | 0.81 | 0.88 | 0.77 |
| Test | 11.46 | 12.55 | 2.82 | 2.64 | 3.19 | 2.23 |

The baseline predictions for renewable energy transition in these three regions, assuming that development is in a steady state and there are no additional energy transition policies, are depicted with dashed blue lines in Fig. 5. We denote this the baseline scenario without any additional policy instruments. The $RNWXYEAR$ of Singapore is forecasted to jump from 2019 to 2020, mainly due to the recent installation of solar panels. After that, the growth rate in the amount of renewable power generation slows due to the declining local economy after COVID-19, although recent studies show that the percentage of renewables out of all power generation may have increased due to an even sharper decline in fossil power generation [59,60]. The forecasted $RNWXYEAR$ is 238.8 gigawatt hours (GWh) in 2025. However, the $RNCAP$ of Singapore is forecasted to be 405.0 megawatt-peak (MWp) in 2025, which is almost double the capacity in 2019. This shows that with existing energy transition policies, renewable infrastructure projects in Singapore are mainly funded by government agencies and are less impacted by short-term economic downturns. In London, a slight decrease in $RNWXYEAR$ is predicted in 2021, but the increasing rate recovers its momentum thereafter and the estimated generation is 1,323 GWh in 2025. Mild but steady growth is expected for $RNCAP$, which ends up at 515.4 MWp in 2025 and is 16.6% higher than the capacity in 2019. In contrast, California is witnessing a major transformation in renewable energy development. Its $RNWXYEAR$ and $RNCAP$ are expected to significantly increase to 74,422 GWh and 35,349 MWp, which translate to an 80% and 98% increase from the levels in 2019, respectively.

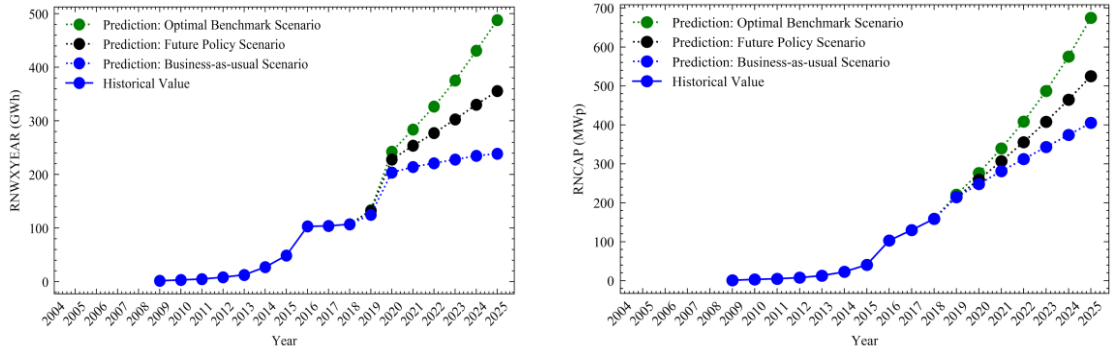

(a) Singapore's *RNWXYEAR* (left) and *RNCAP* (right)

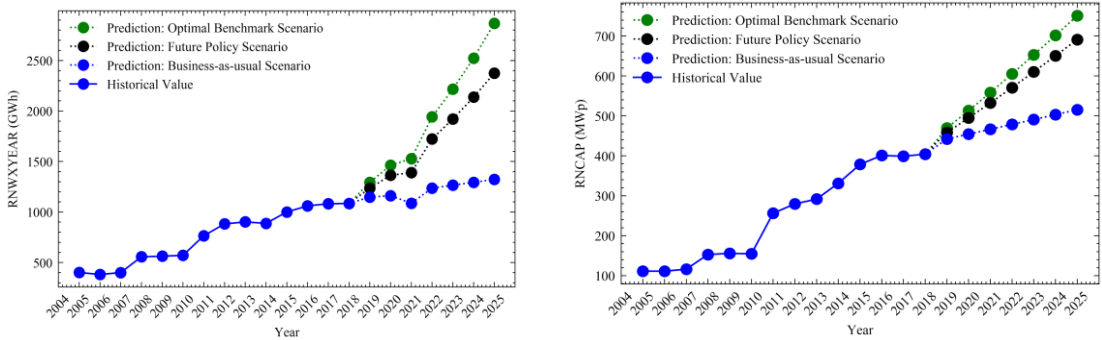

(b) London's *RNWXYEAR* (left) and *RNCAP* (right)

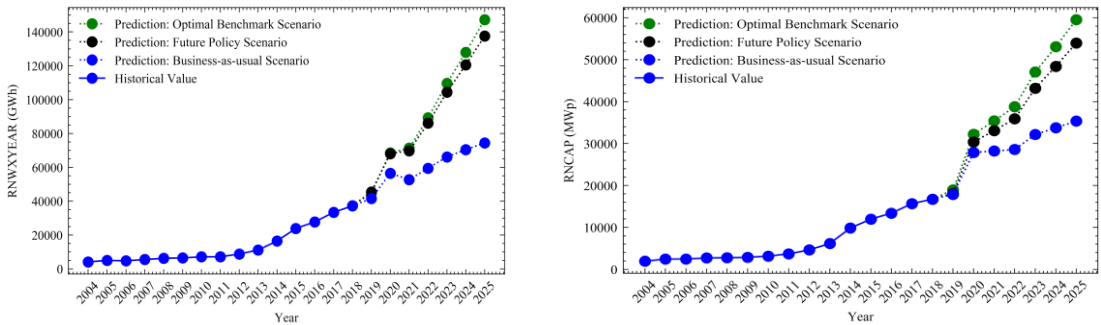

(c) California's *RNWXYEAR* (left) and *RNCAP* (right)

**Fig. 5** Predictions for Energy Transition in (a) Singapore, (b) London and (c) California

Next, we adjust the baseline predictions using different scenarios of policy interventions. We use Denmark as the optimal scenario in our diffusion model of policy impact, since it was the first country to establish a long-term goal of fully replacing fossil energy with renewable energy [61]. According to the latest Danish Integrated National Energy and Climate Plan, all energy consumption in the country is expected to shift to renewables by 2030. Adjusted renewable power generation for the three regions in the optimal scenario (i.e., if the same policy in Denmark were imposed) are plotted with green lines in Fig. 5. In all regions, both the optimal *RNWXYEAR* and *RNCAP* are much larger than the baseline levels without additional policy incentives, which demonstrates the significance of appropriate policy instruments in renewable energy development.

Lastly, to quantitatively estimate the impact of impending policies on renewable power generation in the three regions, we adjust the diffusion ceiling and speed based on region-specific policies by varying the strength of local energy transition policy compared with Denmark's optimal scenario, according to the policy scoring mechanism (see Subsection 2.4). These factors are presented in Table 4. The ceiling factor and growth speed for both production and capacity are smaller in Singapore than in the other two regions, which reflects the differences in progress with respect to the transition to renewable energy in these regions. The predicted *RNWXYEAR* and *RNCAP* are plotted with black lines in Fig. 5. All estimates are smaller than the optimal ones but larger than those at baseline. The difference between blue and black lines denotes the impact of slowing policies on renewable energy production and capacity. It also shows that with the proposed future policies, renewable power generation in California will be the closest to the optimal of the three regions; Singapore has the most potential to increase renewable power generation through policy interventions.

**Table 4** Renewable Energy Diffusion Ceiling and Speed in Singapore, London, and California

| Outcome | Region | Ceiling($c$) | Growth Speed ($p$) |
|---|---|---|---|
| Electricity Production from Renewable Sources (RNWXYEAR) | Singapore | 0.478 | 0.111 |
| | London | 0.800 | 0.139 |
| | California | 0.800 | 0.138 |
| Renewable Power Generation Capacity (RNCAP) | Singapore | 0.478 | 0.082 |
| | London | 0.800 | 0.102 |
| | California | 0.800 | 0.102 |

## 3.3. Policy Adjustment

A prominent feature of our data-driven platform is that it allows government agencies to adjust their energy transition policies in a timely manner based on real-time evaluation results. For example, assume that the Singapore government aims to achieve 450 MWp electricity generation from renewables in 2025. Based on the projected socioeconomic and demographic trends and current policy settings, we expect Singapore to be a bit behind this target, since the predicted *RNWXYEAR* is 405 MWp on our platform. Hence, some adjustment measures can be incorporated into the new policy settings to meet this target. Moreover, it is possible that in 2 years' time, policymakers will re-evaluate policies based on updated economic conditions and find that the theoretically predicted *RNWXYEAR* in 2025 is lower than this target of 400 MWp. In that case, policymakers can adjust their energy transition policies again (e.g., by increasing the amount of financial subsidies) to ensure that the energy transition targets can be achieved on time.

In addition, the model can well incorporate exogeneous disruptions such as recessions and natural disasters. Using the COVID-19 pandemic as an example, regional economic growth or fossil energy consumption is expected to change. Using the updated predictions of these determinants, the model can generate the new forecasts of renewable generation in the post-crisis era. Then, by comparing pre- and post-crisis predictions, policymakers will be able to evaluate and adjust their post-crisis energy transition policies accordingly.

## 4. Practical Implications and Recommendations

## 4.1. Cities/Regions at the Primary Stage of Renewable Energy Development (e.g., Singapore)

Singapore's power generation efficiency—defined as the ratio between electricity production (RNWXYEAR) and generation capacity (RNCAP)—is relatively low compared with London and California (Fig. 6). To promote renewable energy, it is recommended that the government apply more direct and integrated policies to attract more market participants and increase diffusion speed. For power utility firms such as the Singapore Power (SP) Group, since the current proportion of renewable power generation out of total energy generation is very small (around 0.5%), increasing renewable energy utility is unlikely to deter system integration or operations on the existing power grid in the short term. Other cities/regions that, like Singapore, are at the primary stage of renewable energy development are recommended to adopt a similar policy nudge to boost market participation and speed up the diffusion rate.

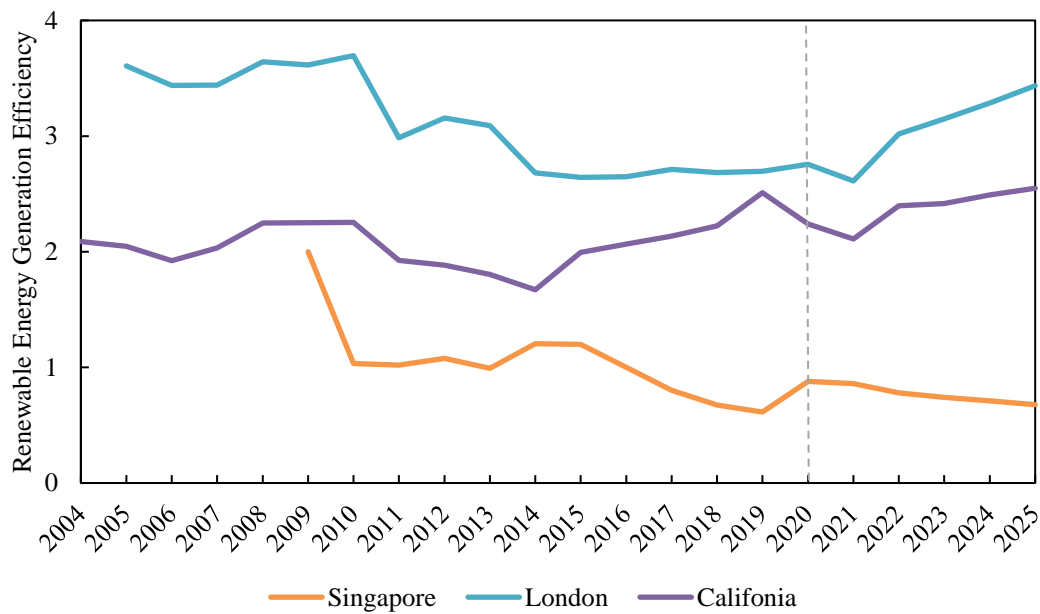

**Fig. 6** Predicted Renewable Power Generation Efficiency in Singapore, London, and California

## 4.2. Cities/Regions at an Intermediate Stage of Renewable Energy Development (e.g., London)

The average growth rate of London's renewable power generation in the next 5 years is predicted to be 2.9% lower than Singapore's (18.7%) and California's (10.4%). As a metropolitan city that adopted renewable energy policies at the earliest stage, London has already established a well-integrated electricity market with both renewable and fossil resources. However, renewable power generation may stagnate in the future, since the estimates predicted under the current policy bundle (in which renewable energy accounts for 6.7% of total generation in 2025) will not be able to meet the city's midterm target (30%). We recommend that London's government agencies introduce more policies to promote local clean energy development. For power utility firms such as British Petroleum (BP), UK Power Networks (UKPN), and National Grid, given that the proportion of renewable power generation is predicted to increase from 3% to 7% soon, the risk of system integration and existing power grid operations should be cautiously evaluated. For consumers and investors, the platform predicts that more feed-in-tariff policies will be provided to accelerate the implementation of renewable power generation facilities and increase diffusion ceilings, which offer opportunities for investment.

## 4.3. Cities/Regions at an Advanced Stage of Renewable Energy Development (e.g., California)

Our evaluation results show that the development of renewable energy in California performs well and is the closest to the optimal scenario in Denmark. California's government agencies should closely monitor the utilisation rate of electricity generated from renewable energy to avoid the risk of oversupply, since overall electricity consumption in the region is predicted to decline in the coming years. In addition to private consumption, better adoption of renewable energy in major public sectors, such as transportation, are also recommended for a sound transition to renewable energy. For electric utilities and system operators, such as California ISO, tremendous growth in renewable power generation is expected to occur in the near future (from 20% to 36% of total generation). We recommend that these entities systematically upgrade the power system for better integration and renew their power scheduling strategies to ensure reliable operations.

## 5. Conclusion

Adjusting energy transition policies appropriately and in a timely manner is challenging but critical, in order to meet growing market demand and engage stakeholders. Most existing policy evaluations are based on retrospective experience, which can be inefficient and generally tackle short-term policy problems. In this paper, we develop a novel data-driven platform to predict and evaluate current renewable energy policies and forecast dynamic pathway to achieve future policy targets, by combining an ANN model and a technology diffusion model. To test the validity of this model, we collect detailed renewable energy policies in three smart regions, Singapore, London, and California, which are at primary, intermediate, and advanced stage of renewable energy development, respectively. By assigning scores on policy effectiveness and combining with rich data on socioeconomic and environmental determinants of renewable energy generation and capacity from 2004 to 2019, we show the powerfulness of this platform in formulating future policy scenarios. Timely adjustment of renewable energy transition policies is crucial in the feedback loop between environmental and socioeconomic driving factors, policy instruments, and renewable energy output.

Specifically, we model different trajectories for renewable energy transitions under baseline, proposed future policy, and optimal scenarios in these three smart regions. Under existing policies, our model predicts that the renewable energy capacity of Singapore will be doubled from 2019 to 2025 with the launch of government-led renewable infrastructure projects, indicating that such policy interventions are relatively resilient to short-term economic fluctuations. The growth of renewable energy capacity in London is expected to be steady and mild, increasing by 17% from 2019 to 2025. In contrast, with the ambitious energy transformation initiatives, the renewable power generation in California is estimated to have a 98% increase in 2025, from the level in 2019. Furthermore, by incorporating different strength of proposed policy interventions into our policy scoring mechanisms, our model generates renewable energy diffusion ceiling parameter (varying from 0.478 to 0.8) and speed parameter (varying from 0.082 to 0.102 in renewable power generation capacity) across the three regions. By comparing future policy scenario with theoretically optimal scenario, we show that with proposed future energy policies, the renewable power generation in California will be the closest to the optimal scenario, while Singapore has the most potential to increase renewable power generation through policy interventions. Thus, a prominent feature of our data-driven platform is to allow government agencies to have timely adjustment of renewable energy transition policies based on real-time evaluation results, which is crucial in the feedback loop between environmental and socioeconomic driving factors, policy instruments, and renewable energy output.

Our results have implications for policymaking in other metropolitan regions. While ambitious policy instruments push regions to the frontier of renewable technologies and speed energy transition, they may be at the risk of discouraging some stakeholders. With the proposed forecasting platform, government agencies can establish realistic and measurable renewable energy transition targets under existing economic conditions and energy market structures. In addition, Policymakers are advised to monitor diffusion speed in renewable energy transition via the proposed forecasting platform and perform timely adjustments based on regulations and incentives. This is especially important when a region's economy or environment undergoes unexpected changes, such as recessions, natural disasters, or pandemics [62,63].

Our research has limitations. Because of data unavailability, we are short of monthly data before 2004 on socioeconomic and environmental determinants of renewable energy generation/capacity. We also lack data at higher frequency to capture more detailed variations in the generation/capacity of renewable energy. Future studies are warranted along the line of incorporating more historical data on socioeconomic and environmental determinants at higher frequency, to improve the model evaluation and forecasting process.

# Appendix. Sources and Predictions of Determinant Variables

**Table A1** Data Source of Determinants

| ID | Name | Abbreviation | Data Source | Link |
|----|------|--------------|-------------|------|
| 1 | Australia Coal Price | COAL | World Bank Open Data | https://thedocs.worldbank.org/en/doc/5d903e848db1d1b83e0ec8f744e55570-0350012021/related/CMO-Historical-Data-Monthly.xlsx |
| 2 | Henry Hub Natural Gas Price | NG | World Bank Open Data | https://thedocs.worldbank.org/en/doc/5d903e848db1d1b83e0ec8f744e55570-0350012021/related/CMO-Historical-Data-Monthly.xlsx |
| 3 | WTI Crude Oil Price | WTI | World Bank Open Data | https://thedocs.worldbank.org/en/doc/5d903e848db1d1b83e0ec8f744e55570-0350012021/related/CMO-Historical-Data-Monthly.xlsx |
| 4 | Gross Domestic Product | GDP | Singapore: Statistics Singapore London: London Datastore California: US Bureau of Economic Analysis; UCLA An'erson's Economic Outlook | Singapore: https://www.singstat.gov.sg/find-data/search-by-theme/economy/national-accounts/latest-data London: https://www.ons.gov.uk/economy/grossdomesticproductgdp/bulletins/regionaleconomicactivitybygrossdomesticproductuk/1998to2019 California: http://www.dof.ca.gov/Forecasting/Economics/Indicators/Gross_State_Product/ |
| 5 | Interest Rate | IR | Singapore: Statistics Singapore London: Bank of England California: Federal Reserve Board of USA | Singapore: https://www.singstat.gov.sg/find-data/search-by-theme/industry/finance-and-insurance/latest-data London: https://www.bankofengland.co.uk/-/media/boe/files/monetary-policy/baserate.xls?la=en&hash=EEB8729ABFFF4B947B85C328340AE5155A99AD0F California: https://www.federalreserve.gov/releases/h15/ |
| 6 | Population | PPL | United Nations: World Population Prospects 2019 Singapore: Statistics Singapore London: London Datastore California: Department of Finance, State of California; Public Policy Institute of California | Singapore: https://www.singstat.gov.sg/find-data/search-by-theme/population/population-and-population-structure/latest-data London: https://www.ons.gov.uk/aboutus/transparencyandgovernance/freedomofinformationfoi/populationoflondon California: https://www.dof.ca.gov/forecasting/demographics/projections/ |
| 7 | Human Developm | HDI | Global Data Lab | https://globaldatalab.org/shdi/shdi/SGP+GBR+USA/?levels=1%2B4&interpolati |
| | ent Index | | | on=0&extrapolation=0&nearest_real=0 |
| 8 | CO2 Emission | CO2 | Singapore: Statistics Singapore London: London Datastore California: California Air Resources Board | Singapore: https://tablebuilder.singstat.gov.sg/table/TS/M891321 London: https://www.gov.uk/government/statistics/uk-local-authority-and-regional-carbon-dioxide-emissions-national-statistics-2005-to-2017 California: https://ww2.arb.ca.gov/ghg-inventory-data |
| 9 | Electricity Consumption | ECON | Singapore: Singapore Energy Market Authority London: London Datastore California: U.S. Energy Information Administration | Singapore: https://www.ema.gov.sg/Statistics.aspx London: https://data.london.gov.uk/dataset/electricity-consumption-borough California: https://www.eia.gov/electricity/data.php |
| 10 | Electricity Price | EP | Singapore: Singapore Energy Market Authority London: UK Office for National Statistics California: U.S. Energy Information Administration | Singapore: https://www.ema.gov.sg/Statistics.aspx London: https://www.ons.gov.uk/economy/inflationandpriceindices/timeseries/l53e/mm23 California: https://www.eia.gov/electricity/data.php |
| 11 | Total Bright Sunshine Duration | SOLAR | Singapore: Singapore National Environment Agency London: UK Met Office California: UCANR | Singapore: https://data.gov.sg/dataset/sunshine-duration-monthly-mean-daily-duration London: https://www.metoffice.gov.uk/pub/data/weather/uk/climate/stationdata/eastbournedata.txt California: http://ipm.ucanr.edu/WEATHER/wxactstnames.html |
| 12 | Wind Speed | WIND | London: Energy Trend: UK Weather California: UCANR (Excluded for Singapore) | London: https://www.metoffice.gov.uk/pub/data/weather/uk/climate/stationdata/eastbournedata.txt California: http://ipm.ucanr.edu/WEATHER/wxactstnames.html |
| 13 | LCOE of Solar PV | SOLARLCOE | Lazard, IRENA | Lazard: https://www.lazard.com/perspective/levelized-cost-of-energy-levelized-cost-of-storage-and-levelized-cost-of-hydrogen/ IRENA: https://www.irena.org/publications/2021/Jun/Renewable-Power-Costs-in-2020 |
| 14 | LCOE of Wind Turbine | WINDLCOE | Lazard, IRENA (Excluded for Singapore) | Lazard: https://www.lazard.com/perspective/levelized-cost-of-energy-levelized-cost-of-storage-and-levelized-cost-of-hydrogen/ IRENA: |

https://www.irena.org/publications/2021/Jun/Renewable-Power-Costs-in-2020

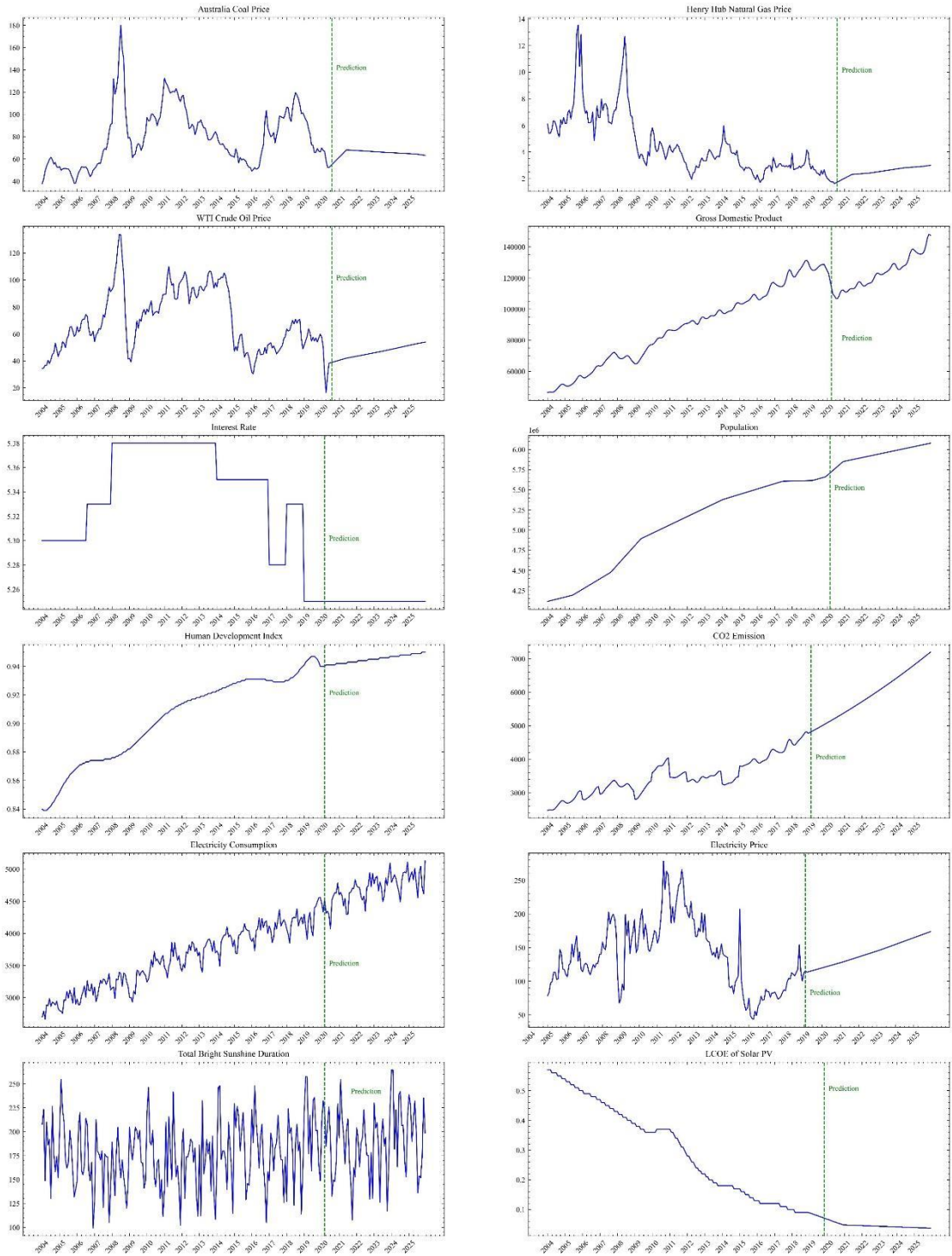

**Fig. A1** Prediction of Determinants in Singapore

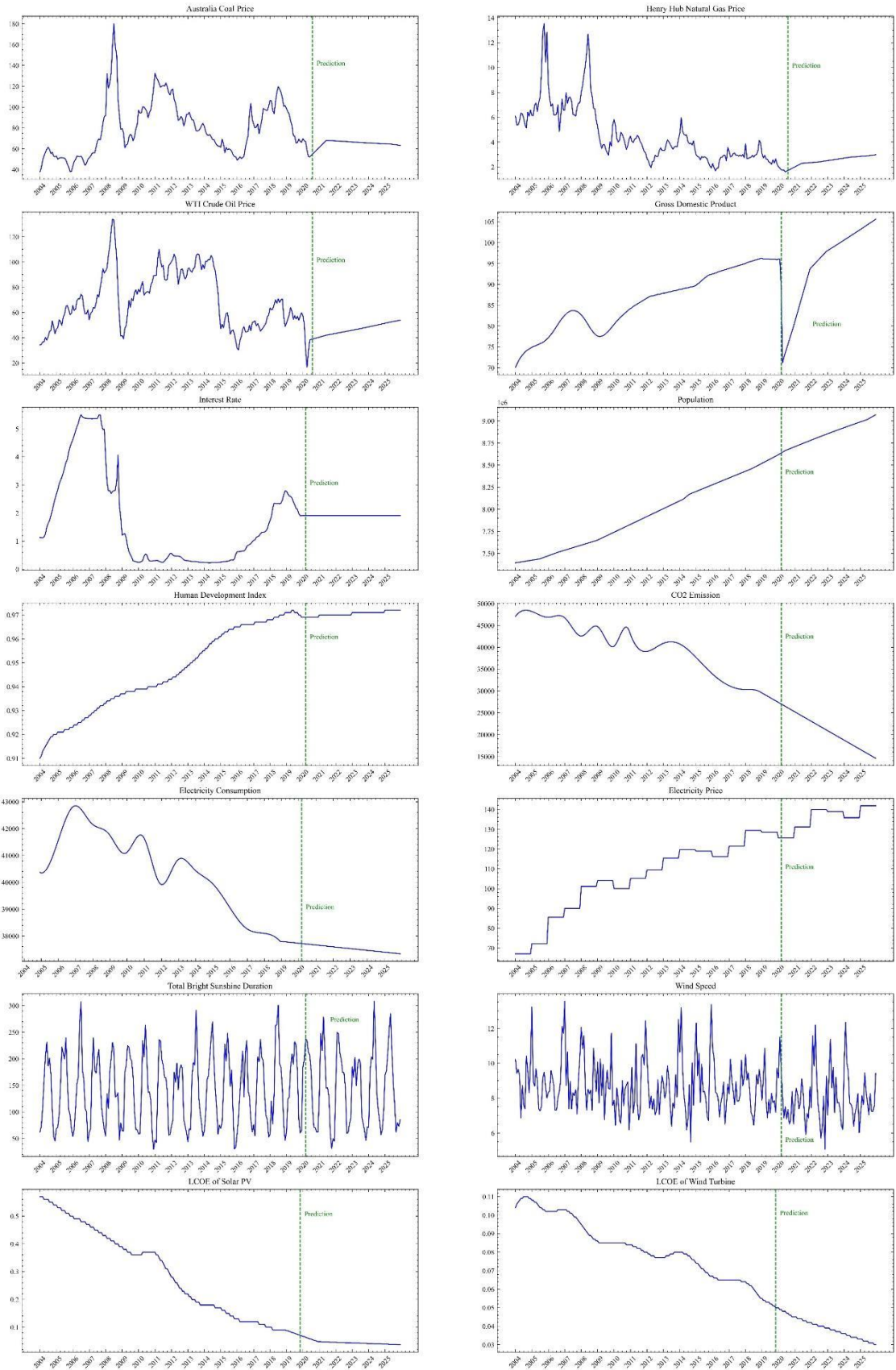

**Fig. A2** Prediction of Determinants in London

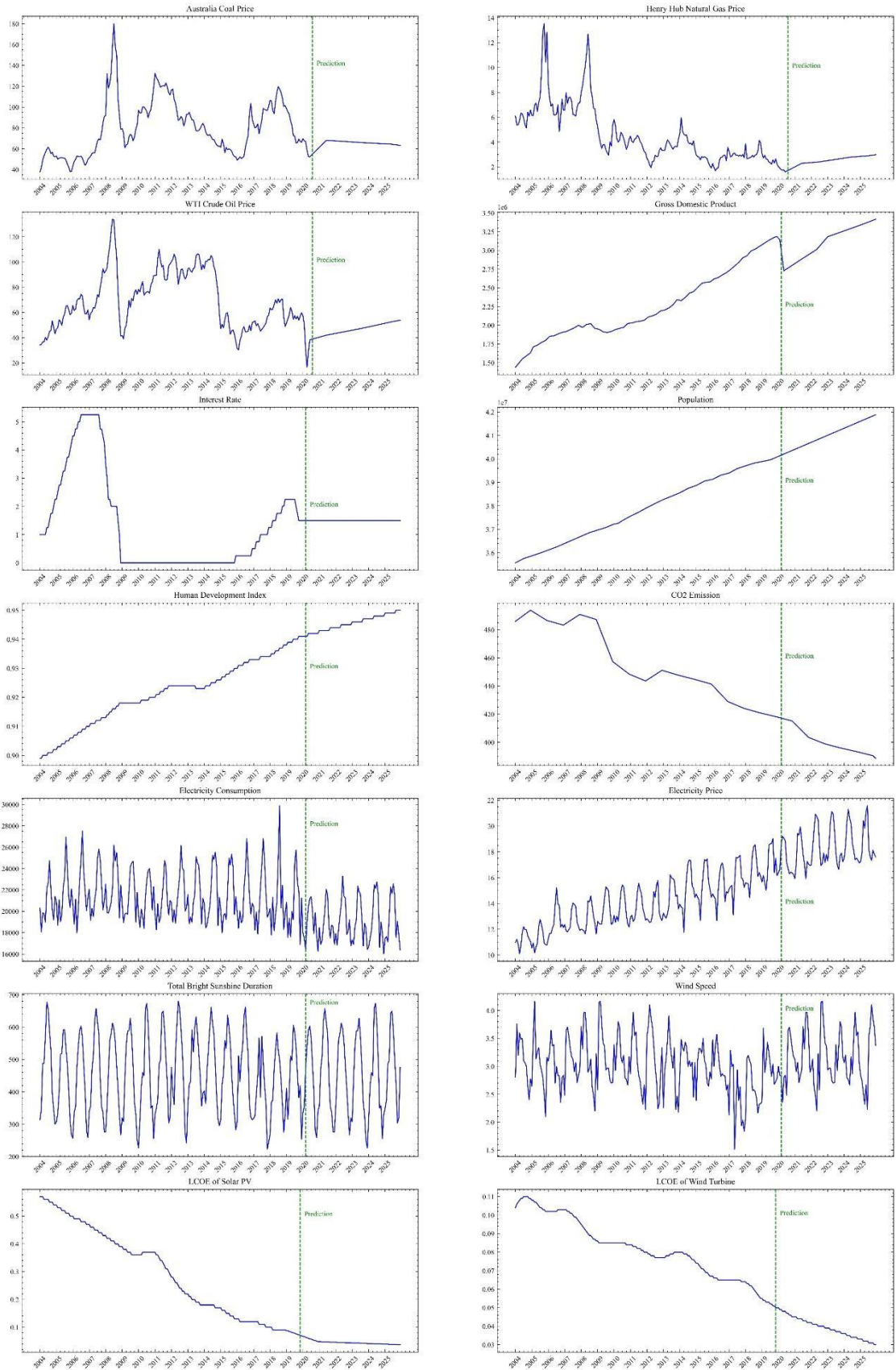

**Fig. A3** Prediction of Determinants in California